# The laminar flow instability criterion and turbulence in pipe

S.L. Arsenjev, I.B. Lozovitski[1], Y.P. Sirik

*Physical-Technical Group*
*Dobroljubova street 2, 29, Pavlograd, Dnepropetrovsk region, 51400 Ukraine*

The identification of stream in the straight pipe as a flexible rod has allowed to present the criterion expression for determination of transition of the laminar flow regime to the turbulent as a loss of stability of the rectilinear static structure of translational motion of stream in pipe and its transition to the flexural-vortical dynamic structure of translational motion, just as a flexible rod buckling. The introduced criterion allows to take into account an influencing of the inlet geometry, the pipe length, the flow velocity, and also of any physical factors on stability of the rectilinear flow structure. It is ascertained, that Reynolds' number is the number of local hydrodynamic similarity and it is displayed that one is constituent part of the introduced stability criterion. The developed approach to a problem of stability is applicable for a problem solving on internal flow and external streamline.

Pacs: 01.65.+g; 46.32.+x; 47.10.+g; 47.15Fe; 47.27Cn; 47.27Wg; 47.60.+i; 51.20.+d; 83.50.Ha

**Introduction**

The fluid has such essential property as an intrastructural mobility in contrast to the solid body. In other words the number of degrees of freedom of the fluid motion is much more, than for the solid body. In this connection the description of the fluid motion is much more complicated than description of the solid body motion. In the beginning we shall view the liquid flow (for example, water), remaining at that within framework of a mechanics. The preservation of volume is naturally for liquid and at the same time the liquid takes any shape, depending on the form of vessel or the flowing element, in which one is contained. The phenomenon of transition from rectilinear structure of laminar translational motion of liquid in pipe to vortical structure of turbulent translational motion, is accompanied with increase of resistance to a motion, was ascertained by experiment.

This is, in basic, traditional information, on the basis of which the attempts of description of the considered phenomenon were undertaken during more than 100 years.

«Many hypotheses are offered for explanation of origin of a turbulence, often very witty hypotheses from a mathematical point of view, however…» L. Prandtl [1].
Till now it is not ascertained:
– What is the turbulence from a physical point of view?
– What is the reason of origin of the turbulence?

The unsolvability of these problems, having the obviously mechanical nature, causes necessity at least of short revision of experience accumulated in a mechanics during last about 400 years.
At the beginning of the XVII century, G.Galilei has established some laws of the free motion of the solid body is thrown by bevel way to horizon in the gravity field. The first principle of mechanics was formulated on the basis of these experiments.

---

[1] Phone: (38 05632) 38892, 40596
E-mail: loz@inbox.ru

In 1641, E.Torricelli has repeated the Galilei's experiments for the liquid free jets. These experiments have shown fundamental uniformity of a free motion of the solid body and the liquid (water) jets in the gravity field. So, the class of problems of mechanics under a title the ballistics was established.

After invention of H.Pitot's instrument (1732), D-I.Bernoulli (1738) has expressed the thought: the correlation of the potential and kinetic energy in the water stream in pipe and a falling body is identical. In the middle of the XIX century this thought was affirmed by J.Weissbach's and H.Darcy's experiment with such feature, that the contact interaction of stream with wall of pipe influences to the correlation of the potential and kinetic energy in stream and to the magnitude of the outflow velocity. Thus, the liquid stream is appeared capable, to a certain extent, to maintain the ballistic properties at motion in pipe.

To the middle of the XIX century, it was also revealed (G.Hagen 1839, 1854), that the disturbance of rectilinear structure of a headway of stream happens during increase of the motion velocity of the water stream in the straight pipe.

To 1895 O.Reynolds has published results of his own experiments on investigation of the transition of motion of the water stream in the straight pipe from rectilinear to vortical structure. Reynolds visualized the flow along its axis by injection of a tinctured jet and he determined, that the flow maintains rectilinear structure of motion at low velocities, and the vortical structure of flow appears at the certain high velocity. To determination of evocative transition, Reynolds offered the dimensionless expression linking the inertial forces of stream with forces of its viscosity. In accordance with Reynolds' expression, the transition of laminar structure of motion of the liquid stream to the turbulent happens by that earlier, than more the velocity of fluid (It means, in a context of our comprehension, the more the intensity of the contact interaction of stream with a pipe wall, the earlier transition happens.) The pipe inlet in the Reynolds' experimental set up is realized in the form of the fluently convergent nozzle. Reynolds did not know apparently, that the inlet convergent mouthpiece can excite the initial turbulence despite of the smoothness of the streamlined contour because of high sensibility of it to an asymmetry of the inleakage. The similar inlet can provoke so-called «bath effect» in the kind of swirl of stream on inlet into pipe and the centrally disposed tinctured jet will not detect the swirl.

In 1929 J.Nikuradse visualized the water stream in a straight chute and shown by means of the filming with cine-camera moved along a stream, that the structure of motion of turbulent flow has the dynamic flexural-vortical character. In 1933 Nikuradse has presented results of his experiments by the water stream in pipes with a different roughness in the form of diagram in coordinates «hydraulic friction coefficient - Reynolds' number». The transition of the laminar flow regime to the turbulent is legibly seen on this diagram. However the problem is remained unclosed: how can engineer use the diagrams of Nikuradse, Colebrook and White, Moody and the other, if the laminar flow regime in pipe can be within the limits $160 \le Re_{cr} \le 40000$?

To investigation of transition of the laminar flow regime to the turbulent, the experiments are being realized, as a rule, with use of long pipes (up to 2000 calibers and more). For example, J.M.T. Thompson has utilized an almost 6-m pipe (more than 910 calibers) even in the home conditions [2]. The next problem is: how does the pipe length influence on transition of the laminar flow regime to turbulent?

Only to the end of the XX century, the guesses about the mechanism of transition of laminar flow in the straight pipe to the turbulent as about phenomenon of a static buckling of the flexible solid body are arised. It was offered to compare the flow stability in the plastic curved pipe with the buckling of cylindrical or spherical shells [2].

Alongside with it, on the other authoritative scientists opinion [3], «there is the large group of problems on stability of the liquid jets. The problem of stability of the water jets (fountain, *notice of the authors*) in air is classical. In particular, what height can be reached by jet if an exit veloc-ity and the jet diameter are assigned?» That is, in the end of the XX century, it is difficult to distinguish: what are the problems of ballistics by Galilei - Torricelli and what are the problems about the contact interaction?

**Approach**

In view of above stated, it is necessary to return to consideration of motion of the free liquid jets in the gravity field. We shall watch the high-viscosity liquid jet, falling vertical, for example, of the melted bitumen at the temperature closely to a solidification point, to accumulative vessel in contrast to Torricelli's and Reynolds' experiments. We can ensure an outflowing jet with the circular cross-section and the flat jet. We can see that the circular cross-section jet deviates from the vertical because of gyration and one is being packed by rings at sufficient height of falling, at contact with the accumulative vessel bottom. The flat jet swings similarly to a pendulum and, being bended, one is being packed up with formation of rouleau of the folds at contact with the above-mentioned vessel bottom. We observe the stability loss phenomenon of foot of the free high-viscous jet similarly to the flexible rod buckling under action of own weight in these experiments. We note the dynamical character of the stability loss phenomenon of jet in contrast from the static buckling of flexible rod. We cannot detect the above-mentioned features at carrying out of the same experiment with the falling jets of water. The distinction in results of the compared experiments can be explained by influencing of viscosity. In this connection, the viscosity is necessary to interpret not as the factor of tractive resistance, but as a physical property of medium to conduct the mechanical energy. Viscosity is fluence of the mechanical energy, spreaded in mediums with the sound speed in accordance with dimensionality – $kgf \cdot m/(m^2 \cdot s)$. The ratio of viscosity to the normal elastic modulus of medium in a direction of action of a mechanical energy determines a relaxation time:

$$t = \frac{\mu}{E} \sim \frac{kgf \cdot s}{m^2} \cdot \frac{m^2}{kgf} = s \,.$$

Having multiplied the sound velocity in the jet material by a relaxation time, we find the height of foot of the free falling jet, which one not discontinuing the flowing, was turned into the flexible rod having the buckling under action of own weight by L.Euler. The free falling jet of the low-viscosity liquid (for example, water) can not make such effect, as a height of action of an braking impulse for its practically is equal to a zero because of its very low viscosity. At the same time the conditions for loss of stability of rectilinear structure of translational motion of stream as the flexible rod are being created at motion of the low-viscous stream (such as water) in the straight pipe. This is stipulated by that the pipe walls, constraining the static head of a stream, transmit its along the stream with increase from the end of stream to its start. The continuity of static head along the pipe length and accordingly along the stream length transmutes the stream into the rod, loaded by the longitudinally allocated forces of the contact interaction with pipe. The straight pipe plays the dual role in the given situation: on the one hand, the pipe guides the stream and shapes the form of the straight rod to it, on the other hand, the pipe loads the stream with longitudinal forces of the contact interaction with it and instigates thereby the loss of stability of rectilinear structure of translational motion of stream as a buckling of the flexible rod under action of own weight by Euler. By the way, Euler reached the correct formulation of problem about the elastic stability of the rod under action of own weight for 34 years (1744-1778) and exact solution of the problem is obtained after 150 years [4].

Thus, the fluid can simultaneously be in state of translational motion and lose the stability of the static rectilinear structure of this motion similarly to the flexible rod, because of the greater number of degrees of freedom as contrast to solid body. At that the dynamic character of loss of stability detected in the above-mentioned experiments with the falling jets of high-viscous liquid is remained at motion of stream of low-viscous fluid in pipe. Moreover, the bending oscillations of stream of low-viscous fluid in pipe lead to the formation of vortices accompanying the translational motion of stream. The described pattern of motion of the turbulent stream quite corresponds to the above-mentioned cinema sequences obtained by Nikuradse.

**Solution**

We imagine the fluid flow core in pipe in the form of the straight flexible rod. The hydrodynamic sublayer coating the roughness of the pipe wall and enveloping the stream core, loads the core by the axial (longitudinal) compressing forces of viscosity, distributed along the pipe length, but the sublayer is remained indefinitely compliant in radial direction and is assumed in the form of the clearance space between the stream core and pipe. We receive that one end of flexible elastic rod, simulating the stream core, is rigidly fastened, and other end has possibility of the axial and - in limits of the clearance radial space - transversal pliability, taking into account the peculiarities of the starting hydrodynamic zone of fluid flow in pipe. We receive simplistically the equality of the stream core diameter to the pipe diameter because of small thickness of hydrodynamic sublayer. The condition of transition of the fluid flow in pipe to turbulent regime consists in equality of the total of the viscosity forces applied to the stream core from the pipe wall side, to Euler's critical force at which one the flexible rod loses the stability of the rectilinear shape of equilibrium under action of the own weight.

The overall value of axial force corresponding to static head of the laminar flow of fluid in the straight pipe with the round cross-section is expressed by J.Poiseuille's formula:

$$N_P = h \cdot \gamma \frac{\pi \cdot D^2}{2 \cdot 4} = \frac{32}{D^2} \mu \cdot L \cdot V \frac{\pi \cdot D^2}{2 \cdot 4} = 4\pi \cdot \mu \cdot L \cdot V, \tag{1}$$

where $h$ – pressure drop, m; $\gamma$ - weight density of fluid, kgf/m$^3$; $D$ – inner diameter of pipe, m; $\mu$ - coefficient of dynamic viscosity, kgf·s / m$^2$; $L$ – pipe length, m; $V$ – average velocity determinated by the flow rate in pipe, m/s.

The minimum quantity of the resultant of the uniformly distributed along length compressing forces, applied to flexible rod, when this rod, having the rigidly fastened end, loses stability by Euler, is determined by expression:

$$N_E \leq \frac{\pi^2 \cdot E \cdot J}{(0.723 \cdot L)^2}, \tag{2}$$

where $E$ – modulus of elasticity, kgf/m$^2$; $J$ – moment of inertia of the flexible rod section, m$^4$.

Euler's stability condition has the appearance for rod with the round section:

$$N_E \leq \frac{\pi^2 \cdot E}{(v \cdot L)^2} \cdot \frac{\pi \cdot D^4}{64}, \tag{3}$$

where $v$ - reduction coefficient of the length of flexible rod, dimensionless quantity.

We obtain the criterial expression for determination of transition of stream in pipe from laminar to turbulent regime, equating forces $N_P$ and $N_E$ under condition of equality of the rod and stream diameters and realizing the elementary transformations:

$$25.9 \frac{\mu}{E} \cdot \frac{V}{L} \left( v \cdot \overline{L}^2 \right)^2 \geq 1, \tag{4}$$

where $\overline{L}$ - relative length of pipe, expressed in calibers of its section, $\overline{L} = L/D$.

Having designated $\mu/E = T_r$ and $L/V = t_r$, we write criterial expression (4) in the form:

$$25.9 \frac{T_r}{t_r} \left( v \cdot \overline{L}^2 \right)^2 \geq 1 \qquad (5)$$

Symbol $T_r$ means the time of mechanical relaxation of fluid by dimensionality and quantity, and $t_r$ is the time of passage of pipe by the stream particles at the obtained expression. Thus, the physical constant of fluid, the generalized geometrical parameter of pipe with allowance for reduction coefficient and the kinematic parameter of stream are introduced into correspondence at the obtained expression.

The analysis of the obtained expressions (4, 5) shows that influence of length of the fluid stream in pipe on the stability of the form of its motion is much higher, than influence of length of flexible rod on its buckling.

The turbulent regime cannot come into being under any increase of velocity in the sufficiently short pipe (nozzle), according to expression (4). Such property of this expression predetermines a presence of the initial zone of stream possessing an absolute stability similarly to the short rods, for which one the problem of stability is not actually. The obtained expression indicates also that the flow will be only laminar at diminution of pipe diameter to the size of capillary, because of the dominant action of the adhesion-cohesion forces of solid surface in this case. Well-known phenomenon of the obliteration stopping the outflow of fluid in the course of time, is natural for such, even of short channels. The obtained criterion expression (4) can be reduced to the form suitable for gas stream and containing the Reynolds' number. It is enough to introduce the substitution $E = p = \frac{1}{k} \cdot a^2 \cdot \rho$ to one and also ones numerator and the denominator to multiply on $V$ and $D$ for this purpose:

$$25.9 \frac{k \cdot \mu \cdot V^2 \cdot D}{a^2 \cdot \rho \cdot V \cdot L \cdot D} \cdot \left( v \cdot \overline{L}^2 \right)^2 = 25.9 \frac{k \cdot M^2}{Re} \cdot \left( v \cdot \overline{L}^{\frac{3}{2}} \right)^2 \geq 1, \qquad (6)$$

where $k$ – adiabatic exponent, $a$ – sound speed, $\rho$ -fluid density, $M$ – Mach number.

The obtained expression (6) is wider than expression for a Reynolds' number at the expense of the more complete and the generalized taking into account of the form of pipe by the non-dimensional parameter, containing not only the diameter of pipe, but also its length with allowance for the length reduction coefficient.

Comparing the transition to the turbulence with buckling of the flexible rod located in pipe with the clearance space, it is necessary to note also difference between them. The stress in such rod is the total of the longitudinal compressing forces and curving forces at its buckling under action of the axial compressing force:

$$\sigma = \frac{P}{A} + \frac{E \cdot J \cdot \Delta \cdot \pi}{l^2 \cdot W}, \qquad (7)$$

where $P$ – axial force of compression, $A$ – sectional area of the rod, $E \cdot J$ – flexural stiffness of the rod, $\Delta$ - clearance space between the rod and the pipe, $l$ – length of quarter of wave of the incurvated rod up to the point of contact with pipe at buckling [5], $W$ – section modulus of the rod.

This stressed state of the rod is reversible as the elastic buckling is considered.

The above-adduced expression (7) to certain extent is valid and for the fluid stream in pipe at the transition to the turbulent flow regime. The difference from behavior of a rod is, that the bending oscillations of the fluid flow core are bound not only with the arcuation of its trajectory of motion, but also with the formation of vortices at periodic pushing of the stream core against the walls of the straight pipe. Both of these kinds of the diversion from rectilinear structure of the translational motion form the nonreversible part of the power losses in turbulent flow. At the same time the direct and reverse transition between the laminar and turbulent regime of a fluid stream in the straight pipe (so called "intermittent turbulence") is characterized by inequality of the pressure drops applied to the pipe flow system in accordance to the experiment results. This feature accompanies also the buckling and the unbending of flexible rod, located into tough pipe with gap and loaded by longitudinal forces of own weight [5].

The features of loss of stability of the laminar flow regime in pipe under action of frictional forces (as well as of flexible rod, under action of own weight) can be traced during viewing of the fluid flow in the light diverged and light converged mouthpieces (nozzles).

Let us compare three flexible rods under action of own weight one: cylindrical and two conical by the elastic stability condition. We accept identical an average diameter and length of all three rods. We accept also identical the elastic modulus, density of material of rods and character of fastening of the bearing end. It is possible to think, that the weight of all three rods is practically identical under the small cone angle, in other words the total force (resultant) of the action of rods to its support is approximately identical.

We are noting that the part of a cylindrical rod near the support is the most critical from a point of view of elastic stability. We are also noting, comparing the viewed rods, that the second rod has the little more section near the support as contrast to the first, and the third rod has the little less section near the support. It is not difficult to establish, taking into account that the ultimate load on Euler is proportional (with other things being equal) to the fourth degree of diameter for rods with the circular cross-section, that the small conicity of the simply supported rods under action of own weight leads to an essential odds in its stability. The similar phenomenons of the protracted laminar flow regime in the light converged mouthpieces and the early transition to the turbulent flow regime in the light diverged mouthpieces are well-known in experimental hydromechanics. In this connection a factor, which takes into account quantity and character of the mouthpiece conicity, is necessary to introduce in the criterion expression (4). In result, the expression (4) will accept a view:

$$25.9 \frac{\mu}{E} \cdot \frac{V}{L} \cdot \left(\frac{d_{ex}}{d_{in}}\right)^4 \left(v \cdot \overline{L}^2\right)^2 \geq 1, \qquad (8)$$

where $d_{ex}$, $d_{in}$ – diameters of the outlet and inlet sections of the conical mouthpieces accordingly.

It is also known in experimental hydrodynamics, that the transition to the turbulent flow regime occurs much earlier in hoselines (for example, rubberized-fabric) than in the metal tough pipes. Reynolds' number can descend up to ~ 150 in hoselines. This feature is bound with the low bending stiffness of hoselines. The factor is necessary to introduce in the expression (4) in the form of ratio of bending stiffnesses of the viewed hoseline and tough pipe, which one was applied in experiments of Nikuradse, Colebrook and White, Moody and other, for taking into account of this feature. Now the obtained criterion expression in the form (8) accept a view:

$$25.9 \frac{\mu}{E} \cdot \frac{V}{L} \cdot \left(\frac{E_s J_s}{E_r J_r}\right)^{\frac{1}{m}} \cdot \left(\frac{d_{ex}}{d_{in}}\right)^4 \cdot \left(v \cdot \overline{L}^2\right)^2 \geq 1, \qquad (9)$$

where $E_sJ_s$, $E_rJ_r$ – bending stiffness of the hoselines and tough pipe accordingly, *1/m* – degree of influence of relative flexibility of pipe on transition to the turbulence.

Influence on system "stream-pipe" of the different factors and actions should and can be taken into account by a similar way in the obtained criterion expression (4):
- the laminarization devices and turbulator in pipe and at inlet to its,
- damping of pipe,
- action of oscillations on the pipe, on the stream in it and on the outflowing jet [4, 6],
- relative increase of the gas stream velocity in the pipe with constant value of sectional area,
- heat exchange of stream with the pipe wall (with allowance for the opposite change of viscosity of liquid and gas under thermal action),
- acceleration of pipe under action of external forces,
- suction or injection of the additional mass flow rate and action of other factors, which one can exert the stabilizing or destabilizing influence on flow regime.

Coefficient $v$ characterizes influence of an inlet profile on the calculated length of pipe and one can be within limits $0.715 \leq v \leq 1.12$. The smaller value will match to the smooth inlet into pipe, the greater value will match to the abrupt inlet into pipe.

The presented analysis and development of physical model of transition of the laminar flow regime to the turbulent in pipe allow to apply its to development of physical models of the mentioned transition for other forms of the internal flows.

We shall consider the problem on the stability of laminar motion of stream in the flat channel in the qualitative aspect. The stream should be presented in the form of the flexible plate located with the backlash in the flat tough channel in this case. It is necessary to note the distinction of the buckling of rods from the buckling of plates from a position of the theory of the elastic static stability of plates. This distinction is, that the elastic stability of the one-piece plate under the compression load appears much above, than the elastic stability of plate composed from the separate rods. It means physically, that the total force of buckling of the rod gang, simulating the plate, is equal to the force of buckling of the one rod, multiplied to the quantity of rods. Such result reflects the independence of the transversal deformation of each rod at its buckling. The rods, constituting the one-piece plate, are bound among themselves and this connection restricts the freedom of their transversal deformation under action of the compressing forces. Therefore the one-piece plate is more steady, than the compounded. Hydrodynamic experiment displays, that the stability of the laminar flow is much higher in the flat channel, than in the ordinary pipes [7]. The replacement of static rectilinear structure of the translational motion to the dynamical flexural-vortical structure of the translational motion of stream happens at the transition of flat flow to turbulent regime. The running bend waves of the stream core make the alternating zones of periodic pushing of its to the flat walls of the channel with formation of vortices and the separation zones from wall. The tractive resistance of the stream is determined by roughness of walls of the flat channel in this case, as this takes place in pipes.

The described approach allows to develop the physical model of loss of stability of laminar flow in the annular gap between two pipes as the elastic cylindrical shell in the rigid annular cavity under action of the axially applied own weight. The stability of such laminar flow appreciably depends on the size of the backlash between pipes enveloping a stream in this case. The stability of the laminar regime is rather higher and it approaches to stability of laminar flow in the flat channel in case of rather small backlash. The stability of laminar flow is reduced up to a level of pipe flowing in case of rather major backlash. It is necessary to take into account an external influence at quantitative estimation of the stability criterion of laminar flow in the flat channel and the annular intertubular backlash as it is shown on an example of pipe flowing.

**Discussion of results**

The stability theory by Euler allows to establish, that the quantity of the ultimate load of flexible rod under action of own weight can vary in the ten times, depending on the conditions of fastening of rod. Results of well-known hydrodynamic experiments [7] also testifies, that the laminar regime can exist at tenfold increase of the stream motion velocity as contrast to value of Reynolds' numbers 2320-4000.

Alongside with it is necessary to take into account, that the action on rod of oscillations of the specified direction and frequencies can also rather appreciably change the quantity of the ultimate load, both in the direction of its increase, and in the direction of decrease. That is, the multiple-factor dependence of stability of flexible elastic rod in static conditions is obvious. The fluid medium has appreciably more number of degrees of freedom of motion because of property of the permanent and continuous intrastructural mobility. The dynamics of fluid appreciably exceeds dynamics of a solid body by variety of combinations of a state and motion and by the multiple-factorness of the influencing actions, thanks to mentioned property. The fluid at motion in pipe has a linear elastics modulus similarly to solid body. This property appears at passage of the flat longitudinal waves of elastic compression-distension in the pipe stream of fluid. Simultaneously fluid has the modulus of the volume elasticity under action of static head. The pipe shapes the fluid stream in the form of lengthy rod. The interaction of stream with the pipe wall loads the fluid rod by forces of longitudinal compression in the form of law of static head [8, 9] that is equivalent to action of own weight to the flexible rod. Thus, the fluid stream is moved by <u>translational</u> in pipe and simultaneously loses the rectilinear structure of laminar motion as the rectilinear shape of an elastic static balance. The combination of motion of the fluid stream in pipe with loss of static stability stipulates appearance of new structure of motion. This structure consists of the running waves of the many times incurvated stream, which at such shape of interaction with wall of pipe generates on the one hand vortices, and on the other hand zones of flow separation from the pipe wall. Running waves of the flexure at pushing of stream under angle of attack on the pipe wall destroy the laminar hydrodynamic sublayer coating the roughness of pipe wall at the developed turbulence. Now tractive resistance of stream in pipe depends in the essential degree on the degree of roughness of the pipe wall in complete correspondence with the Nikuradse's diagram. As to the multiple-factorness of influence of external actions on transition of laminar structure of stream to the turbulent, it is stipulated by multiplicity of motion and state of fluid, specially such thermomechanical active medium as the gas.

The problem arises in connection with the above-mentioned: what is the sense of Reynolds' numbers?

The dimensionless ratio of the inertial forces and the viscosity forces of fluid stream disregarding of any other factors and parameters of "stream-pipe" system, is the number of local hydrodynamic similarity. This number has no indications of the stability criterion of rectilinear structure of flowing.

Elucidation of nature of turbulence and determination of its origin criterion liberate Reynolds' number from unusual function to it and the same time not reduce in slightest degree the significance of this hydrodynamic similarity number in the everything else well-known cases.

**Final remark**

The methodology of approach to the problems of loss of stability of the laminar flow and the structure of the turbulent flow in pipe and resolution of these stated above problems are an example for the solution of similar problems of the internal flows and external streamlines of the different form which one will be enunciated in further publications of the authors.

This work, alongside with previous [8, 9, 10, 11, 12] and future publications is executed initiatively and independently by the scientists of Physical-Technical Group within the framework of development of subject «Physics of motion of liquid and gas» during the last 20 years.

---